\documentclass[twocolumn,prl]{revtex4}%
\pdfoutput=1

\usepackage{epsfig}
\usepackage{graphics}
\begin{document}
\title{Quantum Geometry in the Lab}
\author{Craig Hogan}
\affiliation{Fermilab and  University of Chicago}
\begin{abstract}
Standard particle theory is based on quantized matter embedded in a classical geometry. 
Here, a complementary model is proposed, based on classical matter---  massive bodies, without quantum properties--- embedded in a quantum geometry.   It does not describe elementary particles, but may be a better, fully consistent quantum description for  position states in laboratory-scale systems.  Gravitational theory suggests that the geometrical  quantum system has an information density of about one qubit per Planck  length squared. If so, the model here predicts that the quantum uncertainty of  geometry  creates a new form of   noise in the position of massive bodies, detectable by interferometers.
\end{abstract}

\maketitle


   The    theory of space and time  is  based on the ancient notion that everything that happens,  happens at a definite location.   The classical  name ``geometry'' accurately evokes the  mathematical structure, a map of a smooth surface.  The surface and its behavior do not depend on the map--- geometry is a real, continuous physical structure that does not depend on how it is measured.

Quantum physics posits  a  completely different  model of the world.  At a basic level, it makes no reference to space. It is a theory of systems, and there is, as yet, no quantum theory of the space-time system.  Nothing in the quantum world ever happens at a definite place or time. Locations of particles and interactions are in general indeterminate; system and  measurement cannot be cleanly separated, even in principle. 
In particular, there is no  way to measure the location of a space-time event.

The spooky nonlocality of quantum mechanics--- or the spooky unreality of location---  is vividly demonstrated in real experiments.\cite{Ma2012}  An interaction of a particle in one part of the universe  affects the states of all the other particles.  The idea of locality, so central to the idea of space and time, is simply not a property of quantum reality.

Even so,  physics  achieves a razor-sharp understanding of the  behavior of all known particles and fields, using  a blend of geometrical and quantum ideas, called quantum field theory.  A classical geometry is assumed, with no quantum properties.    Quantum  theory is then applied to modes of space-filling matter fields that are not localized in space and time, usually plane waves extending to infinity. Various states of the waves can have particle-like or wave-like behavior.   

 This approximation  works beautifully, for  practical purposes, in all experiments on particles and fields.  However,  quantum field theory cannot be the whole story. Real, dynamical geometry converses with matter--- in relativity,``space-time tells matter how to move, and matter tells space-time how to curve"--- so at some level, the geometric response to quantum matter must also be a quantum system.

\begin{figure}[b]
 \epsfysize=3in 
\epsfbox{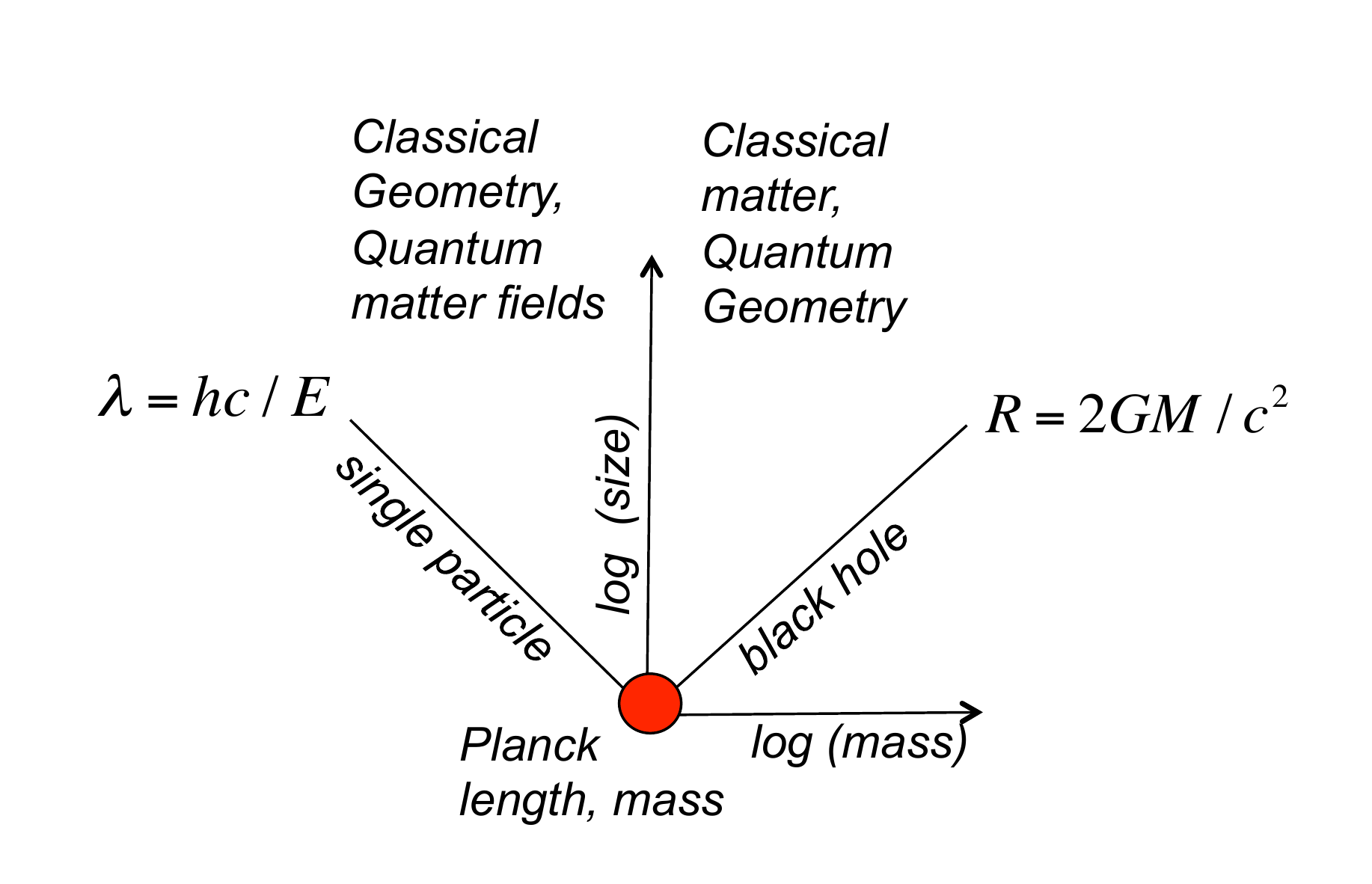} 
\caption{ \label{planckscale}
System size and mass-energy.  The photoelectric relation at left shows the energy of a single  quantum. Smaller systems  do not exist: quanta  do not come in smaller packets of energy. The Schwarzschild formula on the right shows the radius of a black hole. Again, smaller systems do not exist: a black hole is the most compact configuration of space-time for a given energy.  Below the Planck length where the two lines meet,  no system based on classical geometry can exist, suggesting that classical geometry is really an approximate behavior of a quantum system. In the upper region, that system can be approximated (on the left, below the Planck mass) as quantum fields on a classical background, as in standard quantum field theory, or as classical matter on a quantum background, as discussed here.}
\end{figure}

Indeed, below the Planck length, $ct_P\equiv \sqrt{\hbar G/c^3}= 1.616\times 10^{-35}$m, it is no longer consistent to ignore the quantum character of the matter that causes space-time to curve.  Even a single quantum particle of shorter wavelength has more energy than a black hole of the same size, an impossibility in classical relativity
 (see Fig. \ref{planckscale}).  Quantum field theory seems to predict a chaotic ``quantum foam'' of roiling virtual black holes.

Another possibility is that locality,  space-time, and gravity only emerge as an average, approximately classical behavior of a quantum system on large scales.   Such an ``emergent''  space-time could have new fundamental degrees of freedom with a   character different from   ones we know about already, such as gravitational waves or  Standard Model fields.  One can think of analogy with a gas, whose quantum elements, molecules, do not at all resemble  quantized  classical acoustic wave modes; or a solid, whose quantum excitations, phonons, are not  elementary particles.  At the Planck scale, nothing in the   system may  resemble a black hole, a curvature, a metric, or  a position. 
 
A  hint that real space-time is emergent is that the equations of relativity and the principle of equivalence can be derived from a statistical theory,  and equations of motion re-interpreted as thermodynamic relations.\cite{Jacobson:1995ab}  Indeed, even the Newtonian concepts of inertia and gravitational force can be recast in terms of statistics, an ``entropic'' theory of gravity.\cite{Verlinde:2010hp}    We  have not yet identified the quantum states of the geometry, but statistics still predict  classical geometrical behavior, in the same way that the flow of heat and increase of entropy can be understood without knowing  details about atoms.

Although the system  approximates classical space-time on large scales, the geometry may still not behave in exactly a classical way: it may have some quantum indeterminacy even on large scales.   Field theory predicts that such effects are negligible, because effects on the tiny Planck scale average out and do not significantly affect large scale measurements.  But the true quantum geometry may not  separate  scales as well as  field theory does.  

 It is interesting to ask what is predicted in an approximation where classical matter--- massive bodies--- inhabit a quantum geometry, instead of the other way around as in quantum field theory. This way of combining relativity and quantum mechanics complements quantum field theory; it is no good for elementary particles, but may be better for  large  systems, if there really is a quantum geometry.

There are reasons to suspect that quantum geometry may produce new  effects on large scales.
For example, unlike field theory,  emergent gravity implies  a finite amount of geometrical information in any volume, proportional to the area of a bounding surface.  To reproduce classical gravity,  the number of geometrical degrees of freedom in a 3-sphere of radius $R$ should be\cite{Verlinde:2010hp}:
\begin{equation}\label{number}
{\cal N}_{3S}(R)= 4 \pi (R/ct_P)^2.
\end{equation}
An emergent space-time is thus said to be ``holographic''.  It has much less information than standard theory, and does not respect locality; the density and fidelity of spatial information decreases in larger volumes. 

Nobody knows how to create a quantum theory with both  particle and geometrical degrees of freedom, encompassing microscopic to macroscopic scales.  However, it is rather simple to construct  a fully quantum theory for just the geometrical degrees of freedom on large scales,  if we do not  include standard particle modes at the same time.  This macroscopic quantum geometry describes new quantum properties of  collective positions of massive bodies that ordinarily behave in a classical way.

An  effective theory of this kind can be based on a simple noncommutative geometry.\cite{Hogan:2012ne} The mean position of a massive body at rest in 3-space is  described not with classical coordinates, but with quantum operators $\hat x_i$, where $i=1,2,3$. The departure from classical position is described with a noncommutative algebra, 
\begin{equation}\label{3Dcommute}
[\hat x_i,\hat x_j]=  \hat x_k \epsilon_{ijk} ic t_P/\sqrt{4\pi},
\end{equation}
where $\epsilon_{ijk}$ is the antisymmetric tensor. This algebra is well known in the quantum theory of angular momentum, but appears here in a new context, with position in units of the Planck length replacing angular momentum in units of Planck's constant. The eigenstates of the quantum geometrical system form a discrete spectrum, just like components of angular momentum.  The normalization chosen in Eq. (\ref{3Dcommute}) is chosen so that the number of  position eigenstates in a 3-sphere agrees with that required for emergent gravity (Eq. \ref{number}).

Standard quantum mechanics leads to interesting consequences in this new setting.   A radial position operator $\hat L\equiv (\hat x_i\hat x_i)^{1/2}$, like a total angular momentum,  commutes with any position component,  so it  behaves just like a classical separation.
However, like components of angular momentum, the system cannot be a definite position state of two directions at the same time. For bodies separated by distance $L\equiv \langle\hat L\rangle$, the theory  predicts a new quantum-geometrical uncertainty in direction,
\begin{equation}\label{direction}
\langle\Delta \theta^2\rangle  =  ct_P / \sqrt{4\pi}L,
\end{equation}
and in  transverse position, 
\begin{equation}\label{exact}
\langle \hat x_\perp^2 \rangle=   L ct_P/\sqrt{4\pi}= (2.135 \times 10^{-18} {\rm m})^2 (L/{\rm 1 m}).
\end{equation}
Direction and transverse position in this quantum geometry are very slightly indeterminate, even on macroscopic scales.

The illusion of locality--- the appearance, on large scales,  of a property that behaves almost like classical position--- emerges naturally in this theory. The angular uncertainty (Eq. \ref{direction}) becomes smaller on large scales, as the geometrical relationships become ``more classical''. On the other hand, the approach to classical behavior happens more slowly than in quantum field theory.   The transverse uncertainty (Eq. \ref{exact}) actually increases with scale, an effect not present in field theory.  

Actual measurements take random values with a distribution determined by the uncertainty. A system measured over time wanders  in transverse position, with coherent ``motions'' on two-dimensional spacelike sheets defined by light cones around an observer.   Space itself, along with all the matter in it,  appears to jiggle randomly back and forth with amplitude (\ref{exact}) on a timescale $L/c$. The geometry exhibits its own version of quantum weirdness: massive bodies that are close together,  move together, even with no apparent physical connection, since their quantum-geometrical states are entangled by proximity.   The fluctuations, or  ``holographic noise'', are a new effect of quantum geometry; the positional entanglement is the origin of locality.   

The predicted displacement in a laboratory apparatus is many attometers, a detectable distance in an experiment where the massive bodies are mirror elements of an interferometer.\cite{Hogan:2010zs,Hogan:2012ib} The jiggling on a laboratory scale is very slow--- on the order of $10^{-18}c$, comparable with the speed of continental drift--- but with a high frequency,  typically a few Megahertz.
An experiment now being built at Fermilab\cite{holometer} should be capable of either detecting this effect, or conclusively ruling it out. It will  probe the classical coherence of macroscopic quantum geometry  with Planckian sensitivity.

\begin{thebibliography}{}
\bibitem{Ma2012}
X. Ma, S. Zotter, J.  Kofler, R.  Ursin, T.  Jennewein, C.  Brukner, and A. Zeilinger,
Nature Physics,
{\bf 8},
480-485, (2012)



\bibitem{Jacobson:1995ab}
  T.~Jacobson,
  Phys.\ Rev.\ Lett.\  {\bf 75}, 1260 (1995)

\bibitem{Verlinde:2010hp}
  E.~P.~Verlinde,
  JHEP {\bf 1104}, 029 (2011)

  
\bibitem{Hogan:2012ne} 
  C.~J.~Hogan,
  ``Covariant Macroscopic Quantum Geometry,''
  arXiv:1204.5948

\bibitem{Hogan:2010zs} 
  C.~J.~Hogan,
 Phys. Rev. D 85, 064007 (2012)

\bibitem{Hogan:2012ib} 
  C.~Hogan,
  ``Quantum Geometry and Interferometry,''
  in proceedings of the 9th LISA Symposium, Astron. Soc. Pac. Conf. Ser. {\bf 467}, 17 (2012),
  arXiv:1208.3703
  
  \bibitem{holometer}
http://holometer.fnal.gov

\end{thebibliography}
\end{document}